\documentclass[nato,noid,numreferences]{crckbked}

\usepackage{graphicx}
\usepackage{epsf}
\renewcommand{\thesection}{\Roman{section}}
\newcommand{\bea}{\begin{eqnarray}}
\newcommand{\eea}{\end{eqnarray}}
\newcommand{\be}{\begin{equation}}
\newcommand{\ee}{\end{equation}}    
\newcommand{\CA}{A^{(0)}} 
\begin{document}

\begin{opening}
\title{ SPECTRA OF LATTICE DIRAC OPERATORS \protect\\
IN NON-TRIVIAL TOPOLOGY BACKGROUNDS  }

\author{Antonio Gonz\'alez-Arroyo}
\institute{Dpto. de F\'{\i}sica Te\'orica C-XI \\
 and Instituto de F\'{\i}sica Te\'orica UAM-CSIC \\
  Universidad Aut\'onoma de Madrid \\
   Cantoblanco, Madrid 28049, SPAIN}

\begin{abstract}
Dirac operators in non-trivial topology backgrounds in a finite box
are reviewed. 
We analyze how the formalism translates to the lattice, with special 
emphasis on uniform field backgrounds. 
\end{abstract}

\end{opening}

\section{Introduction}

Most of the numerical work performed within Lattice Gauge Theory
makes use of periodic boundary conditions. This has the advantage of
preserving translational invariance and homogeneity of the lattice.
When the box size becomes large in physical units, 
correlation functions become independent of this  size and boundary conditions.
For smaller boxes and/or other observables
the effect of the boundary conditions is non-negligible. This fact has 
to be taken into account when comparing lattice results with the continuum. 

Topological aspects of the gauge fields are directly related to 
boundary conditions. There are well-known difficulties in extending 
topological notions to the lattice. For example, the space of 
three-dimensional (3D) non-abelian gauge 
transformations (or of 4D gauge fields) is
connected on the lattice and disconnected on the continuum. In any case,
it is more appropriate to relate topological properties of the lattice
to those of continuum fields defined on the torus. For example, in trying to
study individual instantons on the lattice there are, not only order $a$
corrections, but also finite size corrections whose origin is in the
continuum: there are no 4D self-dual solutions of topological charge $1$ 
on the torus (without twist).

The topological properties of gauge fields are directly connected with 
the spectrum of the Dirac operator through index theorems. 
Within Lattice Gauge Theories important progress in this respect 
has arisen lately from considerations of 
Ginsparg-Wilson symmetry\cite{GW}, domain-wall fermions\cite{domain} and the overlap 
method\cite{lattchiral}. It is therefore important to re-examine the study of 
the spectrum of lattice Dirac operators for non-trivial topology gauge field 
backgrounds. 
 According to our previous reasoning we should 
consider the case of gauge fields in a box. In a recent paper\cite{uniform} 
the case of uniform (constant) field strength gauge field configurations
in 2 and 4 dimensions has been analyzed in detail (See Ref.~\cite{smitvink} for an early work in this context). In this talk I
will explain and develop  certain aspects with greater depth. 
For simplicity and 
lack of space I  will mostly concentrate in the two-dimensional case.
We refer the reader to Ref.~\cite{uniform} for other aspects not covered in
this paper and a more complete list of references.

\section{Two dimensions: Continuum}

 Charged matter fields on a 2-torus are sections of a U(1) bundle. Without
 sacrificing rigor  we can view them
 as complex  functions $\psi(x)$ defined on the plane and periodic
 up to gauge transformations:
 \bea
 \nonumber
 \psi(x+e_1)&=&\Omega_1(x)\,\psi(x)\\
 \label{BC}
 \psi(x+e_2)&=&\Omega_2(x)\,\psi(x) \quad .
 \eea
 where $e_1=(l_1,0)$ and $e_2=(0,l_2)$ and $l_i$ are the torus lengths. The U(1) fields
$\Omega_i(x)=\exp\{\imath\, \omega_i(x)\}$ are
the transition functions, and must satisfy the following consistency
conditions: 
\be
\label{consistency}
(\omega_1(x+e_2)-\omega_1(x))- (\omega_2(x+e_1)- \omega_2(x))= 2 \pi q\quad .
\ee
where $q$ is an integer characterizing the topology of the bundle. In 
mathematical terms this is the first Chern number of the U(1) bundle. 
The physical interpretation of this integer can be deduced by considering
abelian gauge potentials  on this torus. They must satisfy:
\be
A_j(x+e_i)=A_j(x)+ \partial_j \omega_i(x)
\ee
Now the total flux of the magnetic field $B=\epsilon_{i j} \partial_i A_j$
is given by
\be
\int  dx_1 dx_2\, B = \int dx_2\,  \partial_2\omega_1(x) - \int dx_1\,
\partial_1\omega_2(x)=2 \pi q
\ee
The explicit form of $\omega_i(x)$ is a matter of gauge choice. A
convenient widely used choice is: $\omega_i(x)=\pi q \epsilon_{i j} x_j/l_j$.
A particular
gauge field satisfying these boundary conditions is $\CA_i(x)
=-\frac{F}{2}\epsilon_{i j} x_j$,
where $F=\frac{2 \pi q}{\cal A}$  and ${\cal A}=l_1l_2$ is the area of the torus.
The corresponding magnetic field strength is constant and equal to $F$.
On a torus the  field strength is not the only gauge invariant quantity, one
has also the Polyakov lines winding around non-contractible loops. Hence,
given a pair of real constants $v_i$ (defined modulo $2 \pi/l_i$),
one can construct a whole family of gauge inequivalent gauge potentials 
${A}_i^{(0)\, v}(x)= \CA_i(x) +v_i$ having the same constant field
strength $F$. The  fields $\CA_i$ play an important role in 
parameterizing the space of gauge fields compatible with the boundary
conditions. Indeed, one can make use of Hodge theorem to write an arbitrary 
gauge field as:
\be
\label{Adecomp}
A_i=\CA_i +  v_i + \partial_i \phi + \epsilon_{i j} \partial_j h
\ee
where both $\phi(x)$ and $h(x)$ are real periodic functions on the torus.
We see that $v_i$ parametrises the harmonic 1-forms on the torus. Setting
$\phi=0$ we have the expression of the gauge field in Coulomb gauge.

   Coming back to  matter fields, let us denote by  ${\cal H}$ the space of
functions defined on $\mathbf{R}^2$ satisfying the boundary conditions Eq.~\ref{BC}.
This has the structure of a Hilbert space. Therefore, given a complete set of
commuting operators one obtains an explicit {\em representation} of this space.
Covariant derivatives $D_i=\partial_i-\imath A_i$ with respect to compatible
gauge fields are  anti-hermitian operators defined on ${\cal H}$. 
A natural basis is obtained by selecting the covariant derivatives with 
respect to the uniform field strength potentials:
\be
D_i^{(0)}=\partial_i-\imath \CA_i
\ee
They define  a Heisenberg algebra:
\be
[ D_1^{(0)},  D_2^{(0)} ] =-\imath F
\ee
One can choose (to diagonalize) one the two operators to provide a natural representation
of the space ${\cal H}$. However, for $|q|\ne 1$ this operator alone does
not provide a complete set. There is an additional pair of operators
$K^{(i)}$ commuting with $D_i^{(0)}$ and satisfying:
\be
\label{Heis}
K^{(1)} K^{(2)} = \exp\{\frac{2 \pi \imath}{q}\} \,  K^{(2)} K^{(1)} 
\ee
defining a Heisenberg group. The boundary conditions imply
$(K^{(i)})^q=\mathbf{I}$ and   the space decomposes into a direct sum
of $q$ subspaces. The presence of these $q$ spaces corresponds to the finite
degeneracy of   Landau levels. In infinite space ($q \rightarrow \infty$)
the degeneracy becomes infinite and the relation~(\ref{Heis}) leads to a
new Heisenberg algebra relation. This is the quantum version of the
canonical relations for the Hamiltonian system of a particle in a plane
in an uniform magnetic field ($P_i^{\pm}=p_i\pm \frac{F}{2}\epsilon_{i
j}x_j$ define the two commuting Poisson bracket algebras). 

Our considerations lead to the following parametrization for an arbitrary
element of  ${\cal H}$:
\be
\label{changebasis}
\psi(x_1,x_2)=\exp\{\imath \pi q\frac{x_1 x_2}{{\cal A}}\}\
\sum_{n=1}^q\sum_{s\in{\mathbf Z}}\exp\{2 \pi \imath\frac{x_1}{l_1}(n+sq)\}\
h_n(x_2+\frac{n+qs}{q}l_2)
\ee
where $h_n(y)$ are $q$ arbitrary functions of a single real variable $y$.
This yields a dimensionally reduced description of our Hilbert space. The
price to pay is that, in general, differential operators acting on $\psi(x)$ map onto
integro-differential operators acting on $h_n(y)$. However, covariant derivatives with
respect to gauge potentials depending only on the variable $x_1$ preserve
their differential operator character.

As a particular case one can consider the Dirac operator $D$ in the background
of a uniform magnetic field. The spectrum consists on  q-dimensional spaces of
eigenvectors corresponding to the  eigenvalues $0$ and  $\pm\imath \sqrt{2Fp}$,
where $p$ is a positive integer. We see that the eigenvalues  only depend
on the area of the torus and not on its shape.
The corresponding eigenfunctions in the  $h_n$ representation
are also universal, given in terms of harmonic oscillator eigenfunctions.
Notice, however,  that  shape parameters enter, through  Eq.~\ref{changebasis},
in expressing them in the standard basis. All these results hold for any
value of $v_i$ since (for $q\ne 0$) covariant derivatives 
$\partial_i-\imath {A}_i^{(0)\, v}(x)$ are unitarily equivalent.

\section{Two dimensions: Lattice}    

Now let us examine the situation on a finite $L_1\times L_2$ lattice.
Transition functions do not appear in this formulation, so it seems at
first unclear how one can make contact with the continuum formulation.
However, if we consider compact U(1) lattice gauge fields in the  standard
way ($U_i(n)=\exp\{-\imath A_i\}$),
it turns out that a unique decomposition similar to Eq.~\ref{Adecomp} holds 
on the lattice:
\be
\label{Adeclat}
A_i=\CA_i +  v_i + \Delta^{+}_i \phi + \epsilon_{i j} \Delta^{-}_j h\ \  \bmod
2\pi \mathbf{Z}  
\ee
where $\Delta^{\pm}$ are the forward/backward lattice derivatives, and $v_i$ are
constants defined modulo $2\pi/L_i$. The functions $\phi$ and $h$ are
periodic on the lattice and uniquely defined up to an additive constant.
Finally $\CA_i$ is the lattice version of the constant field strength fields:
\be 
\CA_i= -\frac{F}{2} \epsilon_{i j} I_j(n)
\ee
where $I_1(n)=n_1$ for $1\le n_2< L_2$ and $I_1(n_1,L_2)=(L_2+1)n_1$, and a
similar definition for $I_2$. The lattice  constant field strength 
$F=\frac{2 \pi q}{\cal A}$
with ${\cal A}=L_1 L_2$, contains information of the topological sector $q$.
Apparently we have managed to uniquely define a topology of the lattice
fields. The problem is that the decomposition Eq.~\ref{Adeclat} is singular for fields
having any plaquette on the lattice  equal to $-1$. 

The lattice constant field $\CA_i$  is periodic.
Replacing $I_i$ by $n_i$ would lead  to the naive  non-periodic lattice
discretization of the continuum field. The difference can be interpreted by
considering  that the lattice gauge potential includes the transition function
$w_i(n)=\pi q \epsilon_{i j} n_j/L_j$. 

Lattice fermion fields $\psi(n)$ are elements of a complex vector space of
dimension $2{\cal A}$. Usually one views  these fields  as being
periodic on the lattice.
However,  fermion fields always appear in the action coupled to the gauge
potential  through covariant derivatives. Given the decomposition of the gauge
potentials Eq.~\ref{Adeclat}, one can alternatively consider  the non-periodic
gauge fields  $\CA_i$  explained in the previous paragraph and  non-vanishing 
transition functions. 
In this case, the fermion fields are required to satisfy: 
\be
 \psi(n+e_i)=\exp\{\imath \pi q\, \epsilon_{i j} n_j\}\,\psi(n)  
\ee
where $e_i$ take the same form as in the continuum but expressed in lattice
units (a=1). One might choose an appropriate basis on this space
by requiring that a complete set of commuting operators (matrices) 
are diagonal.
An important class of operators is given by the   covariant shift operators
${\cal T}_i$. All versions of  lattice Dirac operators are elements of the 
algebra generated by these operators and the Dirac matrices.  
As in the
continuum case we will attribute  a special role to the covariant shift operators
for uniform fields ${\cal T}^{(0)}_i$. They satisfy:
\be
\label{TEAlg}
{\cal T}^{(0)}_1 {\cal T}^{(0)}_2 = \xi^{-q}\,  
{\cal T}^{(0)}_2 {\cal T}^{(0)}_1 
\ee
where $\xi=\exp\{\frac{2 \pi \imath }{\cal A}\} $
This relation defines a Heisenberg group. It has been studied in connection
with twist-eaters\cite{te}-\cite{review}. One knows that the group acts 
irreducibly on a vector
space of dimension ${\cal A}/(\gcd(q,{\cal A}))$. This implies that 
for $q=1$ one can fix  a basis of the vector space by diagonalizing any one of the two
 ${\cal T}^{(0)}_i$. If $\gcd(q,{\cal A}) >  1$  there are operators 
commuting with these and allowing to define a complete set
 of commuting operators. If $q$ divides $L_i$ the formulas follow closely the 
 continuum case. There is at least a  q-fold degeneracy of all eigenvalues of 
 any lattice Dirac operator. We refer to \cite{uniform} for details.

 For simplicity, let   us restrict to  the  $q=1$ case. The condition
Eq.~\ref{TEAlg} has to be supplemented with the value of the Casimirs:
\be
\label{CAS}
({\cal T}^{(0)}_i)^{\cal A} = \exp\{-\imath \phi_i\} \, \mathbf{I}
\ee
All pairs of ${\cal A} \times {\cal A}$  matrices satisfying
Eqs.~\ref{TEAlg}-\ref{CAS} are equal up a similarity transformation.
In our case, we have $\phi_i={\cal A} v_i$, so that the spectrum only
depends on $v_i$ up to integer multiples of $2 \pi /{\cal A}$. This
contrasts with the continuum where the spectrum is independent of $v_i$. 

In the basis which diagonalizes ${\cal T}^{(0)}_1$ we have:
\bea
{\cal T}^{(0)}_1&=&\mathbf{Q}=\exp\{-\imath \phi_1/{\cal A}\}\, \mbox{diag}\{1,\xi,\xi^2, \ldots
\xi^{{\cal A}-1}\}\\
{\cal T}^{(0)}_2&=&\mathbf{P}\ \quad \mbox{with} \quad P_{j k}=\delta_{k j+1}\exp\{-\imath \phi_2/{\cal
A}\}
\eea
Thus we see that, irrespectively of the lattice Dirac operator we are using,
the spectrum does not depend on the shape of the box but only on the
area ${\cal A}$ (and $v_i$ mod $2 \pi/{\cal A}$). Indeed, the spectrum also
coincides with that of a twisted reduced model for the group U(${\cal A}$)\cite{tek} at weak coupling.

Now we will investigate the properties of the spectrum of the Wilson-Dirac
hamiltonian $H_{WD}(M,r)=\gamma_3D_{WD}(M,r)$ where
$D_{WD}(M,r)=M+D_N+rW$. The naive lattice Dirac operator $D_N$ and Wilson
term $W$ are given by: 
\bea
D_N=\frac{1}{2} \sum_i \gamma_i({\cal T}_i -{\cal T}_i^\dagger)\\
W=\frac{1}{2} \sum_i (2-({\cal T}_i +{\cal T}_i^\dagger))
\eea
We explicitly indicate the dependence of the Wilson-Dirac  operator 
on the  mass $M$ and the Wilson parameter $r$.

In analyzing  the spectrum of $H_{WD}(M,r)$ it is useful to make use of symmetries.
They must correspond to symmetries of the 
algebra Eq.~\ref{TEAlg}. For example the transformation 
\be
{\cal T}^{(0)}_1 \longrightarrow {\cal T}^{(0)}_2 \ ; \quad {\cal T}^{(0)}_2
\longrightarrow {\cal T}^{(0)\dagger}_1
\ee
preserves Eq.~\ref{TEAlg}. However, it can only be an exact symmetry if it
preserves the Casimirs as well, and this only occurs for $\phi_i=0,\pi$.
In this case, it can be combined with a similar rotation in spin space
($\gamma_i \longrightarrow \epsilon_{i j} \gamma_j$) to produce a unitary matrix
$\mathbf{U}$ commuting with $H_{WD}(M,r)$. The operation $\mathbf{U}$ can 
be interpreted
as a $\pi/2$ rotation of space, which happens to be symmetry even
when $L_1\ne L_2$. Indeed, it generates a finite group of 4 elements
$\mathbf{U}^4=\mathbf{I}$.

For even ${\cal A}$ one can construct  additional symmetry operations associated to
${\cal T}^{(0)}_i \longrightarrow -{\cal T}^{(0)}_i$. Using  them one can show
that $H_{WD}(M,r)$, $H_{WD}(M+4r,-r)$ and $-H_{WD}(-M-4r,r)$ are unitarily 
equivalent. 

In Ref.~\cite{uniform} formulas were given that allow the computation of the
eigenvalues of $H_{WD}(M,r)$ to machine precision. One particularly
interesting aspect is the balance between the number  of positive ($N_+$)
and negative ($N_-$) eigenvalues of  $H_{WD}(M,r)$. Obviously
we have $N_++N_-=2{\cal A}$, while ${\cal I}=N_--{\cal A}$ provides a lattice
definition of the index. It is precisely the index of the Neuberger (overlap)
operator. For large positive or negative values of the mass $M$ it is easy to
show that $N_+=N_-$ and the index vanishes. As we decrease
the mass towards negative values some eigenvalues might move  from positive to
negative or vice versa generating a non-zero index.  Jumps of the index take
place at values of the mass $\bar{M}$ for which there exist $\psi_{\bar{M}}$
satisfying $H_{WD}(\bar{M},r)\psi_{\bar{M}}=0$. Multiplying by $\gamma_3$ the
previous equation, one sees that $\psi_{\bar{M}}$ must be an eigenvector
of real eigenvalue (equal to $-\bar{M}$) of $D_{WD}(0,r)$. Furthermore,
the sign of $\psi_{\bar{M}}^\dagger \gamma_3 \psi_{\bar{M}}$ determines whether
the index increases or decreases at this point. Thus, to analyze the index
of $H_{WD}(M,r)$ it is enough to determine the real eigenvalues of
 $D_{WD}(0,r)$ and the corresponding eigenvectors. For $r=1$ one can apply
 similar techniques to those used in Ref.~\cite{uniform} to derive expressions
 that allow the computation of this real eigenvalues to machine precision.
 For ${\cal A} \ge 3$ there are always four real eigenvalues, two in the
 interval $[-2,0)$ and two in the interval $(-4,-2]$ (For even ${\cal A}$
 there is symmetry around $M=-2$). As we decrease the value of $M$ from
 positive values the index follows the sequence $0\rightarrow 1\rightarrow 0
 \rightarrow -1 \rightarrow 0 $. The physical region at which the index takes
 its continuum value  (equal to one) has an upper edge given by
  $-0.63397, -0.58578,-0.29449,
 -0.12210,  -0.0311755$ for ${\cal A}=3,4,10,25,100$. For large torus sizes
 it approaches $-Fr/2$. The lower edge is  $-0.94203$ for  ${\cal A}=3$ and
decreases very fast towards $-2$ as the  area increases.  One can monitor 
the $M$ dependence of the smallest (in absolute value) eigenvalue of 
 $H_{WD}(M,r)$. The interval $(-4,0)$ is split into 4 regions, one for each of the eigenvectors of real eigenvalue  $\psi_{\bar{M}}$. Within each region 
the behaviour is essentially linear (in $M$) with a slope determined by  
$\psi_{\bar{M}}^\dagger \gamma_3 \psi_{\bar{M}}$.

Other aspects are studied in  Ref.~\cite{uniform}. 
For example, we numerically computed the eigenvalues of
Neuberger's operator in a uniform background gauge field, 
for several torus sizes up to $24\times24$ and
several values of $M$. Eigenvalues and eigenvectors are   quite close 
to the continuum results.

\section{Four dimensions}
Much of the previous construction generalizes to the case of non-abelian
gauge groups in 4 dimensions. For SU(N) gauge groups the topology of 
gauge fields is indexed  by the instanton number $Q$ (and twist sectors). 
To make connection with the two-dimensional U(1) formulation of the previous
sections, we make use of the fact that (except for SU(2) and odd $Q$)
it is possible to construct  transition functions
living in the maximal abelian subgroup U(1)$^{N-1}$\cite{vanbaal1}-\cite{review}. 
Then the boundary conditions for the fermion fields apply  for each 
of the color components independently. 
This reduces the problem from SU(N)  to a U(1) field in 4 dimensions. The
fluxes of these U(1) fields over the 2-planes encodes the information of 
the topology of the original field. Choosing appropriate transition functions
and a basis of the  4 dimensional space, the problem  decouples into two 
 2-planes.  This allows
to obtain the spectrum of the continuum Dirac operator and the lattice naive 
Dirac operator in terms of the two-dimensional result. The Wilson-Dirac and
Neuberger operators, however, couple the 4 dimensions in a more complicated
way. We refer to Ref.~\cite{uniform} for details.

\section{Acknowledgments}
I want to thank the organizers of the workshop for their invitation and 
for creating  a very  stimulating and pleasant atmosphere during the 
workshop. Many of the ideas expressed in this paper originated or
were enriched in exchanges and discussions with my collaborators Leonardo Giusti,
Christian Hoelbling, Herbert Neuberger and Claudio Rebbi. 
I also want to thank Robert Kirchner for some programming help.

  
 \end{document}